# Optical surface waves supported and controlled by thermal waves

Yaroslav V. Kartashov,[1] Victor A. Vysloukh,[2] and Lluis Torner[1]

[1]*ICFO-Institut de Ciencies Fotoniques, and Universitat Politecnica de Catalunya, Mediterranean Technology Park, 08860 Castelldefels (Barcelona), Spain*
[2]*Departamento de Fisica y Matematicas, Universidad de las Americas – Puebla, Santa Catarina Martir, 72820, Puebla, Mexico*

We show the formation of optical surface waves at the very edge of semiconductor materials illuminated by modulated light beams that generate thermal waves rapidly fading in the bulk material. We find families of thresholdless surface waves which existing due to the combined action of thermally-induced refractive index modulations and instantaneous Kerr-type nonlinearity.

Surface waves localized at the interface of two optical materials are intriguing objects, which are promising for sensing, evanescent-wave spectroscopy, surface characterization, or optical switching [1,2]. Nonlinear self-confinement of light beams near the edge of $Al_xGa_{1-x}As$ waveguide arrays with focusing nonlinearity leads to the formation of surface solitons [3,4], and interfaces between defocusing lattices and uniform media support surface gap solitons [5-8]. Surface lattice solitons in two-dimensional geometries have been also observed [9, 10]. The surface-wave localization is made possible by a nonlinear contribution to the refractive index that shifts the propagation constant into the forbidden gaps of the periodic lattice spectrum.

  A challenging open problem is the exploration of the opportunities afforded by dynamical refractive index modulations. Induction of such landscapes by thermal waves [11] sets a particularly fascinating example. Thermal waves are excited upon illumination of the material edge with time-modulated light beams. It has been shown experimentally that focused laser beams heats surface up to $\sim 100$ °C [12]. In combination with large, positive thermo-optic coefficients of such materials as $Al_xGa_{1-x}As$ ($dn/dT \simeq 2.67 \times 10^{-4}$ $K^{-1}$ [13]), heating results in significant refractive index changes.

In this Letter we introduce transient optical surface waves supported by the refractive index landscapes induced by diffusive thermal waves. We discuss the salient features of thresholdless surface waves propagating in the transparency band of the material. Such waves exist due to the jointed action of the thermally-induced refractive index profile and Kerr-type nonlinearity. Importantly, we show that the near-surface solitons might be steered by acting on the parameters of illumination that induces the thermal wave.

We consider semiconductor plate in the coordinate plane $(\eta, \xi)$, which is heated along its left edge $\eta = 0$ by a homogeneous (along $\xi$), harmonically-modulated light wave $I_0[1+\cos(\Omega t)]/2$, where $I_0$ is the intensity and $\Omega = 2\pi/\tau$ is the frequency. In the visible wavelength range light is strongly absorbed [14] thus generating heat. The right plate edge $\eta = L$ is assumed to be thermo-stabilized at the temperature $T_r$, the ambient temperature is $T_l$, while the upper and lower surfaces of plate are thermo-isolated. The temperature distribution along $\eta$-axis is governed by the equation:

$$\frac{\partial T}{\partial t} - \frac{\partial^2 T}{\partial \eta^2} = \frac{1}{2}\exp(-\alpha\eta)[1+\cos(\Omega t)], \qquad (1)$$

with the boundary conditions $\partial T/\partial \eta |_{\eta=0} = \sigma(T-T_l)$, and $T|_{\eta=L} = T_r$. Here $\sigma$ is the coefficient of thermal losses, and $\alpha$ is the absorption coefficient. The coordinate $\eta$ is normalized to the transverse scale $x_0$; time $t$ is expressed in units of $t_s = x_0^2/\kappa$, where $\kappa$ is the thermo-diffusion coefficient; the temperature is normalized to $T_s = \alpha I_0 x_0^2/(\rho C_p)$, where $\rho$ is the density of the material, and $C_p$ is its thermal capacity. Under experimental conditions, one has $\rho = 5$ g/cm$^3$, $C_p = 0.33$ J/(gK), $\kappa \simeq 0.1$ cm$^2$/s, while $x_0 = 10$ $\mu$m corresponds to $t_s \sim 10^{-5}$ s.

The absorbed radiation acts as a distributed heat source with a penetration depth $\alpha^{-1}$. It creates a steady-state temperature profile $T_{st}(\eta)$, governed by the equation $-\partial^2 T_{st}/\partial \eta^2 = \exp(-\alpha\eta)/2$ with the boundary conditions $\partial T_{st}/\partial \eta |_{\eta=0} = \sigma(T_{st}-T_l)$, $T_{st}|_{\eta=L} = T_r$. In practically interesting case of strong absorption and small thermal losses ($\alpha \gg 1$, $\sigma \ll 1$) one gets $T_{st} \simeq T_m - (T_m - T_r)\eta/L$, where $T_m$ is the surface tempera-ture. This steady-state temperature profile produces an almost constant refractive index gradient that deflects light beam to the left or to the right edge of the sample, depending of the sign of $T_\mathrm{m} - T_\mathrm{r}$.

Dynamical thermal wave profiles are governed by the equation $\partial T_\mathrm{tr}/\partial t - \partial^2 T_\mathrm{tr}/\partial \eta^2 = \exp(-\alpha\eta)\cos(\Omega t)/2$ with time-varying heat source, and boundary conditions $\partial T_\mathrm{tr}/\partial \eta|_{\eta=0} = \sigma T_\mathrm{tr}$ and $T_\mathrm{tr}|_{\eta=L}=0$. We are interested in steady-state, temporally-periodic solutions of this equation. The analytical solutions are represented by a cumbersome infinite series, so we proceed directly with the numerical solution easily obtained a finite-differences method. Typical snapshots of the temperature distributions formed when transients dissipate (further we omit subscript "tr") are shown in Fig. 1(a) for different instances at $\tau=16$, $\alpha=10$, $\sigma=0.006$, $T_\mathrm{l}=0$, $L=9\pi$. These profiles are well approximated by $T(\eta,t) \simeq T_\mathrm{m}\exp(-\gamma\eta)\cos(\Omega t - \gamma\eta - \pi/4)$, where the fading rate $\gamma \sim \tau^{-1/2}$. Figure 1(b) shows the thermal wave profiles for different modulation periods at $t=\tau/8$, when the temperature reaches its maximum at the sample edge. One can control the penetration depth $\gamma^{-1} \sim \tau^{1/2}$ by varying the modulation period $\tau$. Note that a sample with a width $L \gg \gamma^{-1}$ might be considered semi-infinite, and the duration of the transition process scales as the square of the sample width.

Figure 1(c) illustrates the dependence of the maximal transient temperature $T_\mathrm{m}$ at $\eta=0$ on the modulation period $\tau$. This temperature monotonically growths with $\tau$ approximately as $T_\mathrm{m} \sim \tau^{1/2}$, it can be changed in a wide range by varying the intensity of heating wave. Since the Eq. (1) is linear, one can combine heating sources with different peak intensities $I_m$, modulation frequencies $\Omega_m = m\Omega$ (here $m$ is a positive integer), and phases $\varphi_m$ [i.e., to use pulsed thermal sources $\sim \sum_{m=1}^{M} I_m \cos(\Omega_m t + \varphi_m)$], to generate thermal waves with a variety of complex shapes. Figure 1(d) shows example of three-hump thermal waves excited at $t=0$ by three heat sources having different parameters. The possibilities of thermal wave shaping are even richer, since one can heat the sample with light generated by lasers with different wavelengths having remarkably different absorption lengths [14].

We now consider the propagation of a laser beam along the interface, inside the time-periodic refractive index landscape created by the thermal waves via the thermo-optic effect. We assume that the wavelength of the propagating light beam belongs to the transparency band ($\lambda=1.53\,\mu\mathrm{m}$ for $\mathrm{Al_xGa_{1-x}As}$), so that it is not absorbed even though its intensity is high enough to produce refractive index changes due to Kerr-type nonlinearity [4]. The light propagation is described by the nonlinear Schrödinger equation:

$$i\frac{\partial q}{\partial \xi} = -\frac{1}{2}\frac{\partial^2 q}{\partial \eta^2} - |q|^2 q - pT(\eta,t)q, \qquad (2)$$

where $\xi$ is the longitudinal coordinate normalized to the diffraction length $L_{\mathrm{dif}} = kx_0^2$; $k = 2\pi n_0/\lambda$ is the wavenumber; the parameter $p = L_{\mathrm{dif}}^2(dn/dT)\alpha I_0/(n_0\kappa\rho C_{\mathrm{p}})$ characterizes the depth of the refractive index profile induced by $T(\eta,t)$; $n_0$ is the unperturbed refractive index; and $q$ is the normalized complex amplitude. We searched for stationary solutions of Eq. (2) (parametrically depending on time) in the form $q(\eta,\xi) = w(\eta)\exp(ib\xi)$, where $w$ is a real function and $b$ is the propagation constant. The area $\eta < 0$ is supposed to be occupied by a linear medium with $n_0 \simeq 1$, that leads to the total internal reflection at $\eta = 0$.

Thermal increase of the refractive index may give rise to formation of linear surface waves. Figure 2 illustrates the properties of such linear optical modes. Their propagation constants are presented in Fig. 2(a) as a function of the refractive index modulation depth $p$. The plot contains information about the cutoffs for higher-order modes that require a certain minimal modulation depth $p$ for their existence. Figure 2(b) shows the normalized profiles of linear modes.

Figure 3 illustrates the possibility of time-dependent steering of near-surface solitons by thermal waves. We integrated the nonlinear Eq. (2) with input condition $q(\eta,0) = \mathrm{sech}(\eta - \eta_0)$ at $\eta_0 = 4$, and the soliton beam propagates almost undisturbed in the absence of thermal wave. The thermal wave induces refractive index gradient that deflects this soliton beam to the right or to the left, depending on time [see Fig. 1(a)]. Figure 3(a) shows intensity distributions for the case of strongest soliton deflection to the left ($t = \tau/8$), followed by reflection, and strongest deflection to the right ($t = 5\tau/8$). The dependence of the output soliton center displacement on time is non-monotonic, especially at large distances, when reflection from the interface occurs [Fig. 3(b)]. The scanning speed is around few kHz for a soliton width of the order of $x_0 = 10\ \mu\mathrm{m}$, and the output soliton displacements substantially exceed a soliton width.

Thermally induced refractive index landscapes can also support transient nonlinear surface modes. The fundamental surface mode becomes narrower with growth of $b$, while its intensity maximum shifts to the surface [Fig. 4(a)]. The energy flow $U = \int_0^\infty w(\eta)^2 d\eta$ of the fundamental mode is a monotonically growing function of $b$ [Fig. 4(b)]. We found also a variety of higher-order modes [Fig. 4(c)] which bifurcate from the linear surface waves [see Fig. 2(b)] in low-energy limit, and, hence, all of them are thresholdless. Fundamental surface modes are linearly stable in the entire domain of their existence. A detailed linear stability analysis shows that higher-order modes can also be stable, except for a narrow region close to cutoff for existence on $b$ [Fig. 4(d)]. We found that the width of the instability domain decreases with growth of $p$ and that above a critical propagation constant, higher-order surface waves do become stable.

Summarizing, we have shown that thermal waves excited at the edge of suitable nonlinear materials, like semiconductors, support thresholdless optical surface waves, and afford time-dependent steering of solitons propagating near the surface. Importantly, a rich variety of dynamical refractive index landscapes may be generated by acting on the depths, shapes, and transverse extents of the thermal waves, a feature that opens up the possibilities for light manipulation and control.

# References with titles

1. *Nonlinear surface electromagnetic phenomena*, Ed. by H. E. Ponath and G. I. Stegeman, North Holland, Amsterdam (1991).

2. D. Mihalache, M. Bertolotti, and C. Sibilia, "Nonlinear wave propagation in planar structures," Progr. Opt. **27**, 229 (1989).

3. K. G. Makris, S. Suntsov, D. N. Christodoulides, and G. I. Stegeman, "Discrete surface solitons," Opt. Lett. **30**, 2466 (2005).

4. S. Suntsov, K. G. Makris, D. N. Christodoulides, G. I. Stegeman, A. Haché, R. Morandotti, H. Yang, G. Salamo, and M. Sorel, "Observation of discrete surface solitons," Phys. Rev. Lett. **96**, 063901 (2006).

5. Y. V. Kartashov, V. A. Vysloukh, and L. Torner, "Surface gap solitons," Phys. Rev. Lett. **96**, 073901 (2006).

6. M. I. Molina, R. A. Vicencio, and Y. S. Kivshar, "Discrete solitons and nonlinear surface modes in semi-infinite waveguide arrays," Opt. Lett. **31**, 1693 (2006).

7. C. R. Rosberg, D. N. Neshev, W. Krolikowski, A. Mitchell, R. A. Vicencio, M. I. Molina, and Y. S. Kivshar, "Observation of surface gap solitons in semi-infinite waveguide arrays," Phys. Rev. Lett. **97**, 083901 (2006).

8. E. Smirnov, M. Stepic, C. E. Rüter, D. Kip, and V. Shandarov, "Observation of staggered surface solitary waves in one-dimensional waveguide arrays," Opt. Lett. **31**, 2338 (2006).

9. X. Wang, A. Bezryadina, Z. Chen, K. G. Makris, D. N. Christodoulides, and G. I. Stegeman, "Observation of two-dimensional surface solitons," Phys. Rev. Lett. **98**, 123903 (2007).

10. A. Szameit, Y. V. Kartashov, F. Dreisow, T. Pertsch, S. Nolte, A. Tünnermann, and L. Torner, "Observation of two-dimensional surface solitons in asymmetric waveguide arrays," Phys. Rev. Lett. **98**, 173903 (2007).

11. *Photoacoustic and thermal wave phenomena in semiconductors*, Ed. By A. Mandelis, North Holland, New York (1987).

12. P. S. Dobal, H. D. Bist, S. K. Mehta, and R. K. Jain, "Laser heating and photoluminescence in GaAs and $Al_xGa_{1-x}As$," Appl. Phys. Lett. **65**, 2469 (1994).

13. F. G. Della Corte, G. Cocorullo, M. Iodice, and I. Rendina, "Temperature dependence of the thermo-optic coefficient of InP, GaAs, and SiC from room temperature to 600 K at the wavelength of 1.5 $\mu$m," Appl. Phys. Lett. **77**, 1614 (2000).

14. D. E. Aspnes, S. M. Celso, and R. A. Logan, "Optical properties of $Al_xGa_{1-x}As$," J. Appl. Phys. **60**, 754 (1986).

# References without titles

1. *Nonlinear surface electromagnetic phenomena*, Ed. by H. E. Ponath and G. I. Stegeman, North Holland, Amsterdam (1991).
2. D. Mihalache, M. Bertolotti, and C. Sibilia, Progr. Opt. **27**, 229 (1989).
3. K. G. Makris, S. Suntsov, D. N. Christodoulides, and G. I. Stegeman, Opt. Lett. **30**, 2466 (2005).
4. S. Suntsov, K. G. Makris, D. N. Christodoulides, G. I. Stegeman, A. Haché, R. Morandotti, H. Yang, G. Salamo, and M. Sorel, Phys. Rev. Lett. **96**, 063901 (2006).
5. Y. V. Kartashov, V. A. Vysloukh, and L. Torner, Phys. Rev. Lett. **96**, 073901 (2006).
6. M. I. Molina, R. A. Vicencio, and Y. S. Kivshar, Opt. Lett. **31**, 1693 (2006).
7. C. R. Rosberg, D. N. Neshev, W. Krolikowski, A. Mitchell, R. A. Vicencio, M. I. Molina, and Y. S. Kivshar, Phys. Rev. Lett. **97**, 083901 (2006).
8. E. Smirnov, M. Stepic, C. E. Rüter, D. Kip, and V. Shandarov, Opt. Lett. **31**, 2338 (2006).
9. X. Wang, A. Bezryadina, Z. Chen, K. G. Makris, D. N. Christodoulides, and G. I. Stegeman, Phys. Rev. Lett. **98**, 123903 (2007).
10. A. Szameit, Y. V. Kartashov, F. Dreisow, T. Pertsch, S. Nolte, A. Tünnermann, and L. Torner, Phys. Rev. Lett. **98**, 173903 (2007).
11. *Photoacoustic and thermal wave phenomena in semiconductors*, Ed. By A. Mandelis, North Holland, New York (1987).
12. P. S. Dobal, H. D. Bist, S. K. Mehta, and R. K. Jain, Appl. Phys. Lett. **65**, 2469 (1994).
13. F. G. Della Corte, G. Cocorullo, M. Iodice, and I. Rendina, Appl. Phys. Lett. **77**, 1614 (2000).
14. D. E. Aspnes, S. M. Celso, and R. A. Logan, J. Appl. Phys. **60**, 754 (1986).

# Figure captions

Figure 1. Spatial profiles of thermal waves (a) for different moments of time $t$ at $\tau=16$ and (b) for different modulation periods $\tau$ at $t=\tau/8$. (c) Maximal surface temperature as a function of modulation period $\tau$ at $t=\tau/8$. (d) Complex spatial profile of surface wave created by three heat sources with the parameters $(\tau_1=160,\ \varphi_1=3\pi/4,\ I_1=0.04)$, $(\tau_2=16,\ \varphi_2=5\pi/4,\ I_2=0.16)$, and $(\tau_3=1,\ \varphi_3=3\pi/4,\ I_3=0.80)$ at $t=0$.

Figure 2. Linear surface modes supported by the refractive index profile induced by thermal wave at $t=\tau/8$, $\tau=16$. (a) Propagation constants versus modulation depth $p$. (b) Profiles of first three linear modes at $p=60$ corresponding to points marked by circles in (a).

Figure 3. Steering of near-surface soliton $q(\eta,0)=\operatorname{sech}(\eta-\eta_0)$ by thermal wave at $\eta_0=4$, $\tau=64$, $p=5.5$. Panel (a) shows superimposed field modulus distributions for solitons in the time moments $t_1=\tau/8$ and $t_2=5\tau/8$. (b) Soliton center displacements versus time at different propagation distances.

Figure 4. Nonlinear surface modes supported by thermally induced refractive index profiles at $t=\tau/8$ and $\tau=16$. (a) Profiles of fundamental modes with different $b$ at $p=20$. (b) Energy flow versus $b$ for fundamental and dipole modes at $p=20$. Points marked by circles correspond to profiles shown in (a). (c) Profiles of dipole and triple-mode solutions at $b=4$, $p=60$. (d) Real part of perturbation growth rate for triple-mode solution at $p=60$.

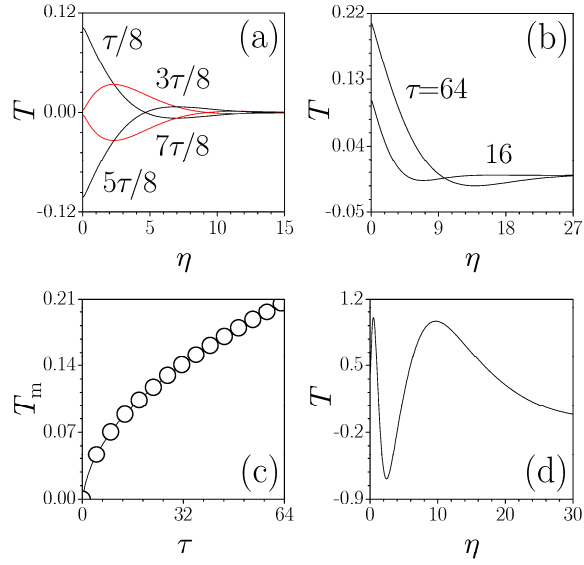

Figure 1. Spatial profiles of thermal waves (a) for different moments of time $t$ at $\tau = 16$ and (b) for different modulation periods $\tau$ at $t = \tau/8$. (c) Maximal surface temperature as a function of modulation period $\tau$ at $t = \tau/8$. (d) Complex spatial profile of surface wave created by three heat sources with the parameters ($\tau_1 = 160$, $\varphi_1 = 3\pi/4$, $I_1 = 0.04$), ($\tau_2 = 16$, $\varphi_2 = 5\pi/4$, $I_2 = 0.16$), and ($\tau_3 = 1$, $\varphi_3 = 3\pi/4$, $I_3 = 0.80$) at $t = 0$.

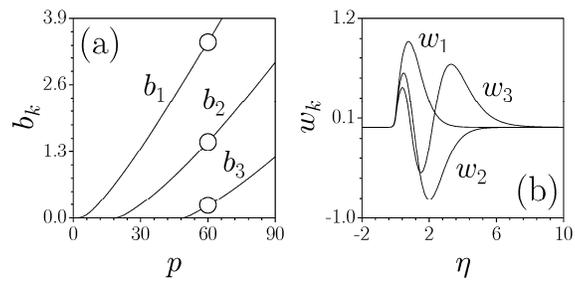

Figure 2. Linear surface modes supported by the refractive index profile induced by thermal wave at $t = \tau/8$, $\tau = 16$. (a) Propagation constants versus modulation depth $p$. (b) Profiles of first three linear modes at $p = 60$ corresponding to points marked by circles in (a).

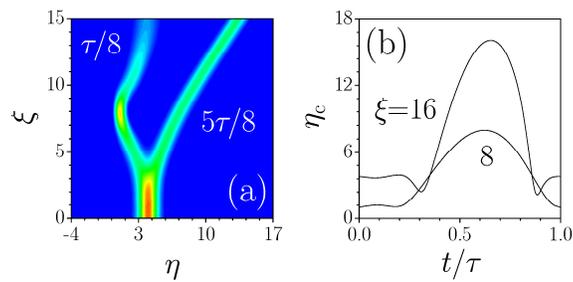

Figure 3. Steering of near-surface soliton $q(\eta,0) = \text{sech}(\eta - \eta_0)$ by thermal wave at $\eta_0 = 4$, $\tau = 64$, $p = 5.5$. Panel (a) shows superimposed field modulus distributions for solitons in the time moments $t_1 = \tau/8$ and $t_2 = 5\tau/8$. (b) Soliton center displacements versus time at different propagation distances.

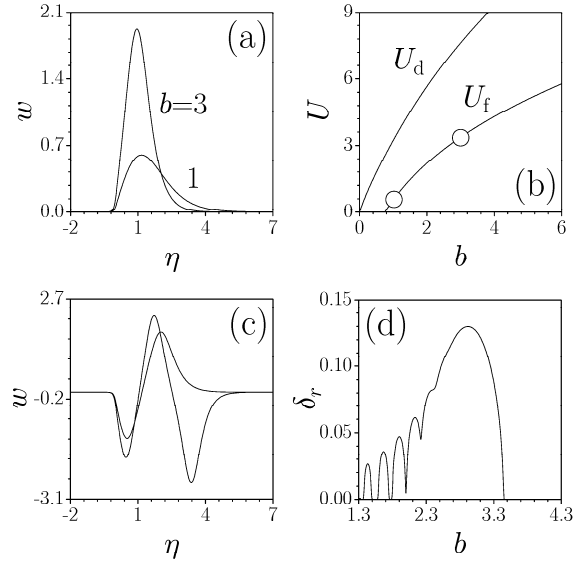

Figure 4. Nonlinear surface modes supported by thermally induced refractive index profiles at $t = \tau/8$ and $\tau = 16$. (a) Profiles of fundamental modes with different $b$ at $p = 20$. (b) Energy flow versus $b$ for fundamental and dipole modes at $p = 20$. Points marked by circles correspond to profiles shown in (a). (c) Profiles of dipole and triple-mode solutions at $b = 4$, $p = 60$. (d) Real part of perturbation growth rate for triple-mode solution at $p = 60$.